\documentclass[preprint]{raa}            
\usepackage{graphicx,times}             
\usepackage{natbib}
\usepackage{amssymb,amsmath}
\bibpunct{(}{)}{;}{a}{}{,}

\usepackage[a4paper=true,dvipdfm=true,pagebackref=true]{hyperref}
\hypersetup{colorlinks = true, linkcolor = green, anchorcolor = red, citecolor = blue, filecolor = red, pagecolor = red, urlcolor = red}

\begin{document}

   \title{Accretion Properties of MAXI J1813-095 during its Failed Outburst in 2018
}

   \volnopage{Vol.0 (20xx) No.0, 000--000}      
   \setcounter{page}{1}          

   \author{Arghajit Jana
      \inst{1}
   \and Gaurava K. Jaisawal
      \inst{2}
   \and Sachindra Naik
      \inst{1}
   \and Neeraj Kumari
      \inst{1,4}
   \and Debjit Chatterjee
      \inst{5,3}
   \and Kaushik Chatterjee
      \inst{3}
   \and Riya Bhowmick
      \inst{3}
   \and Sandip K. Chakrabarti
      \inst{3}
   \and Hsiang-Kuang Chang
      \inst{6}
   \and Dipak Debnath
      \inst{3}
   }

   \institute{Astronomy \& Astrophysics Dividsion, Physical Research Laboratory, Navrangpura, 
   Ahmedabad 380009, India; {\it argha0004@gmail.com}\\
        \and
             National Space Institute, Technical University of Denmark, Elektrovej, 327-328, DK-2800 Lyngby, Denmark\\
       \and
             High Energy Astrophysics, Indian Centre for Space Physics, Garia St Road, Kolkata 700084, India\\
        \and
             Department of Physics, Indian Institute of Technology, Gandhinagar, 382355, Gujarat, India\\
        \and
            Indian Institute of Astrophysics, Koramangala, Bangalore, 560034, India\\
        \and
            Department of Physics, National Tsing Hua University, Hsinchu 30013, Taiwan \\
\vs\no
   {\small Received~~20xx month day; accepted~~20xx~~month day}}

\abstract{ We present the results obtained from a detailed timing and spectral studies of a black hole candidate 
MAXI~J1813-095 using {\it Swift}, {\it NICER}, and {\it NuSTAR} observations during its 2018 outburst. 
The timing behaviour of the source is mainly studied by using {\it NICER} light curves in a $0.5-10$~keV 
range. We did not find any signature of quasi-periodic oscillations in the power density spectra of the
source. We carry out spectral analysis with a combined disk blackbody \& power law model, and physical
two-component advective flow (TCAF) model. From the combined {\tt disk blackbody} \& {\tt power-law} 
model, we extracted thermal and non-thermal fluxes, photon index, and inner disk temperature. We also 
find evidence for weak reflection in the spectra. We have tested physical TCAF model on a broadband 
spectrum from {\it NuSTAR} and {\it Swift}/XRT. The parameters like mass accretion rates, the size of 
the Compton clouds and the shock strength are extracted. Our result show that the source remained in 
the hard state during the entire outburst which indicates of a `failed' outburst. We estimate the mass 
of the black hole as $7.4 \pm 1.5$ $M_{\sun}$ from the spectral study with the TCAF model. We use 
{\tt LAOR} model for the Fe K$\alpha$ line emission. From this, the spin parameter of the black hole 
is estimated as $a^* > 0.76$. The inclination angle of the system is estimated to be in the range of
$28^{\circ} - 45^{\circ}$ from reflection model. We estimate the source distance to be $\sim 6$~kpc.
}

\authorrunning{Jana et al. }   
\titlerunning{Accretion Dynamics of MAXI~J1813-095} 

   \maketitle

\section{Introduction}           
\label{sect:intro}
Transient black hole X-ray binaries (BHXRBs) occasionally show outbursts that last from weeks to months. During the outburst, the X-ray intensity of the source rises thousands of times as compared to that during the quiescent state. An outburst is believed to be triggered when the viscosity is suddenly enhanced at the pileup radius \citep{C90,C96,CND19,Bhowmick20}. A transient BHXRB is known to show characteristic evolution in the spectral and timing properties during these outbursts that are broadly classified as Type~1 and Type~2 outbursts \citep{DD17}. In case of Type~1 outbursts, BHXRBs show all the usual spectral states viz. hard, hard-intermediate, soft-intermediate and soft states due to which these outbursts are called as a full or complete outbursts. On the other hand, in case of Type 2 outbursts, which are also known as failed outbursts, only harder spectral states (hard and hard-intermediate) are observed \citep{Delsanto2016,Tetarenko2016,Garcia2019}.

In general, the spectrum of a BHXRB can be modelled with a power law (PL) continuum model along with a thermal multicolour disk blackbody (DBB) component. In addition, an Fe K$\alpha$ emission line at around $\sim 6.4$~keV is observed \citep{RM06}. It is believed that the DBB component originates from a standard geometrically thin accretion disk \citep{SS73} whereas the power-law component originates from a Compton cloud that consists of hot electrons \citep{ST80}. The soft photons from the standard accretion disk are inverse Comptonized at the Compton cloud and produce a hard power-law tail. Several theoretical models have been developed in the literature to explain the nature of the Compton cloud \citep{Z93,HM93,Esin97}.

The Two-Component Advective Flow (TCAF) model is a generalized accretion flow solution where the transonic flow includes rotation, viscosity and radiative transfer. It can explain the observed spectral and timing properties of BHXRBs self-consistently \citep{CT95,C97}. In this model, the accretion flow has two components: high viscous Keplerian disk with high angular momentum, and low viscous sub-Keplerian halo which has low angular momentum. The Keplerian disk is submerged within the sub-Keplerian flow and moves slowly in the equatorial plane. The sub-Keplerian flow forms an axisymmetric shock at the centrifugal boundary. The post-shock region consists of hot electrons and known as CENBOL or CENtrifugal pressure supported BOundary Layer \citep{C96}. The CENBOL acts as a Compton cloud. The soft photons originate from the Keplerian disk and contribute to the multicolour blackbody component. A fraction of the soft photons are intercepted by the CENBOL, and get inverse-Comptonized and become hard photons that form the hard power-law tail. A fraction of the hard photons is reprocessed at the Keplerian disk and a `reflection hump' is observed at higher energy. In TCAF paradigm, quasi-periodic oscillations (QPOs) which are observed in power density spectra (PDS) are produced by oscillation of the CENBOL \citep{MSC96}. The CENBOL is also the launch site of the jet. In recent years, the TCAF model has been used successfully used to study the spectral and timing properties of several black holes and active galactic nuclei \citep{SM16,DC16,Shang19,Nandi2019,KC2020,Anuvab20}.

Black Hole candidate (BHC) MAXI~J1813-095 was discovered on 2018 February 19 \citep{Kawase18} with the {\it MAXI}/GSC. Follow up observations with the {\it Swift}/XRT localized the source at RA = 18$^h$ 13$^m$ 34.0$^s$, Dec = -09$^{\circ}$ 31$'$ 59$''$.0 \citep{Kennea18}. The GROND observation of the above location detected the optical counter part of the source \citep{Rau18}. The ATCA observation revealed a compact jet and classified the source as the radio-quiet BHXRB \citep{Russell18}. From the multi-wavelength observations, \citet{AP19} suggested that the companion star could be a G5V star with a distance of $>3$~kpc.

In this paper, we studied MAXI~J1813-095 in broad energy bands using {\it Swift}/XRT, {\it NICER}, and {\it NuSTAR} observations performed during the outburst. The paper is organized in the following way. In \S 2, we describe the observation and data analysis processes. In \S 3, we present the timing and spectral analysis results. In \S 4, we discuss our findings and finally, in \S 5, we summarize our results.

\section{Observation \& Data Reduction}

\subsection*{{\it NuSTAR}}
The transient black hole candidate MAXI~J1813-095 was observed with {\it NuSTAR} at three epochs during the declining phase of the 2018 outburst (see Table~\ref{tab:log}). {\it NuSTAR} \citep{Harrison2013} is the first hard X-ray focusing observatory launched by NASA. It consists of two identical focusing modules: FPMA and FPMB. These modules are sensitive to X-ray photons in the range of $3-79$~keV. We reprocessed data from the {\it NuSTAR} observations with the help of {\it NuSTAR} data analysis software ({\tt nustardas}, version 1.4.1\footnote{\url{https://heasarc.gsfc.nasa.gov/docs/nustar/analysis/}}). Cleaned event files were produced and calibrated using standard filtering criteria with the {\tt nupipeline} task by using the latest calibration files\footnote{\url{http://heasarc.gsfc.nasa.gov/FTP/caldb/data/nustar/fpm/}}. We chose a circular region of radii 120 arcsec centered at the source coordinates for the source and away from the source for the background products. The {\tt nuproduct} task was used to extract source and background spectra. We re-binned the source spectra as 20 counts per bin with {\tt grppha}\footnote{\url{https://heasarc.gsfc.nasa.gov/ftools/caldb/help/grppha.txt}} task. 

\subsection*{{\it Swift}}

{\it Swift} observed MAXI~J1813-095 at two epochs simultaneously with two {\it NuSTAR} observations. In total, {\it Swift} observed MAXI~J1813-095 twelve times between 2018 February 20 and 2018 March 25. All the observations were carried out with the {\it Swift}/XRT in the energy range of $0.5-10$~keV. {\it Swift}/XRT observations of MAXI~J1813-095 were carried out in windowed-timing (WT) mode except the first observation which was in photon-counting (PC) mode. We extracted cleaned event files with the {\tt FTOOLS} task {\tt xrtpipeline}\footnote{\url{https://www.swift.ac.uk/analysis/xrt/}}. We choose a circular region of radius 30 arcsec for source and background products. Light curves, source and background spectra were extracted by using {\tt XSELECT} v2.4\footnote{\url{https://www.swift.ac.uk/analysis/xrt/xselect.php}}.

\subsection*{{\it NICER}}

MAXI~J1813-095 was also observed with {\it NICER} at several epochs during the 2018 X-ray outburst. {\it NICER} is an X-ray timing instrument \citep[XTI;][]{Gendreau2012} attached to the International Space Station in 2017 June. It is sensitive to soft X-ray photons in $0.2-12$~keV range. The XTI consists of 56 X-ray concentrated optics, each attached to a silicon drift detector \citep{Prigozhin2012}. There are only 52 detector units active, providing a total effective area of 1900 cm$^2$ at 1.5 keV. An unprecedented timing and spectral sensitivities of $\sim100~ns$ (rms) and $\sim 85$~eV at $1$~keV can also be achieved by {\it NICER}, respectively. For the study of outburst evolution of MAXI~J1813-095, we used publicly available data from {\it NICER} monitoring between 2018 February 21 to February 27. The total effective exposure of these observations with observation ids 1200090101--1200090105 is about 5.5 ks. For analysis, the data were first reprocessed with `{\tt nicerl2}'\footnote{\url{https://heasarc.gsfc.nasa.gov/docs/nicer/analysis_threads/nicerl2/}} script in the presence of latest updated calibration files of version 20200722. Standard GTI was also generated using the `{\tt nimaketime}' task. The cleaned events obtained after the reprocessing were then used for extracting the light curve and spectrum in the {\tt FTOOLS} {\tt XSELECT} environment. For spectral analysis, ancillary response file and response matrix file of version 20200722 are considered in our analysis. The background corresponding to each observation id is simulated by using the {\tt nibackgen3C50}\footnote{\url{https://heasarc.gsfc.nasa.gov/docs/nicer/tools/nicer_bkg_est_tools.html}} tool (Remillard et al, in prep.).

\begin{table}
\begin{center}
\caption{Log of {\it NICER},  {\it NuSTAR}, and {\it Swift}~ observations of the transient black hole candidate
MAXI~J1813-095.}
\label{tab:log}
\begin{tabular}{lccccc}
\hline
ID&Date of Obs. & Obs. ID& Exp \\
& (yyyy-mm-dd) & & (ks)\\
\hline
 &         &{\it NICER}     & \\
X1 & 2018-02-21 & 1200090101 & 0.6  \\
X2 & 2018-02-22 & 1200090102 & 0.4  \\
X3 & 2018-02-23 & 1200090103 & 2.4  \\
X4 & 2018-02-26 & 1200090104 & 1.1  \\
X5 & 2018-02-27 & 1200090105 & 1.3  \\
\hline
&       & {\it NuSTAR} \\
N1 & 2018-02-28 & 80402303002 & 23.2  \\
N2 & 2018-03-06 & 80402303004 & 20.5  \\
N3 & 2018-03-25 & 80402303006 & 20.4  \\
\hline
&       & {\it Swift/XRT}\\
S1      & 2018-03-06 & 00088654002 & 1.8 \\
S2      & 2018-03-25 & 00088654004 & 1.9 \\
\hline
\end{tabular}
\end{center}
\end{table}

\begin{figure}
\includegraphics[width=\columnwidth]{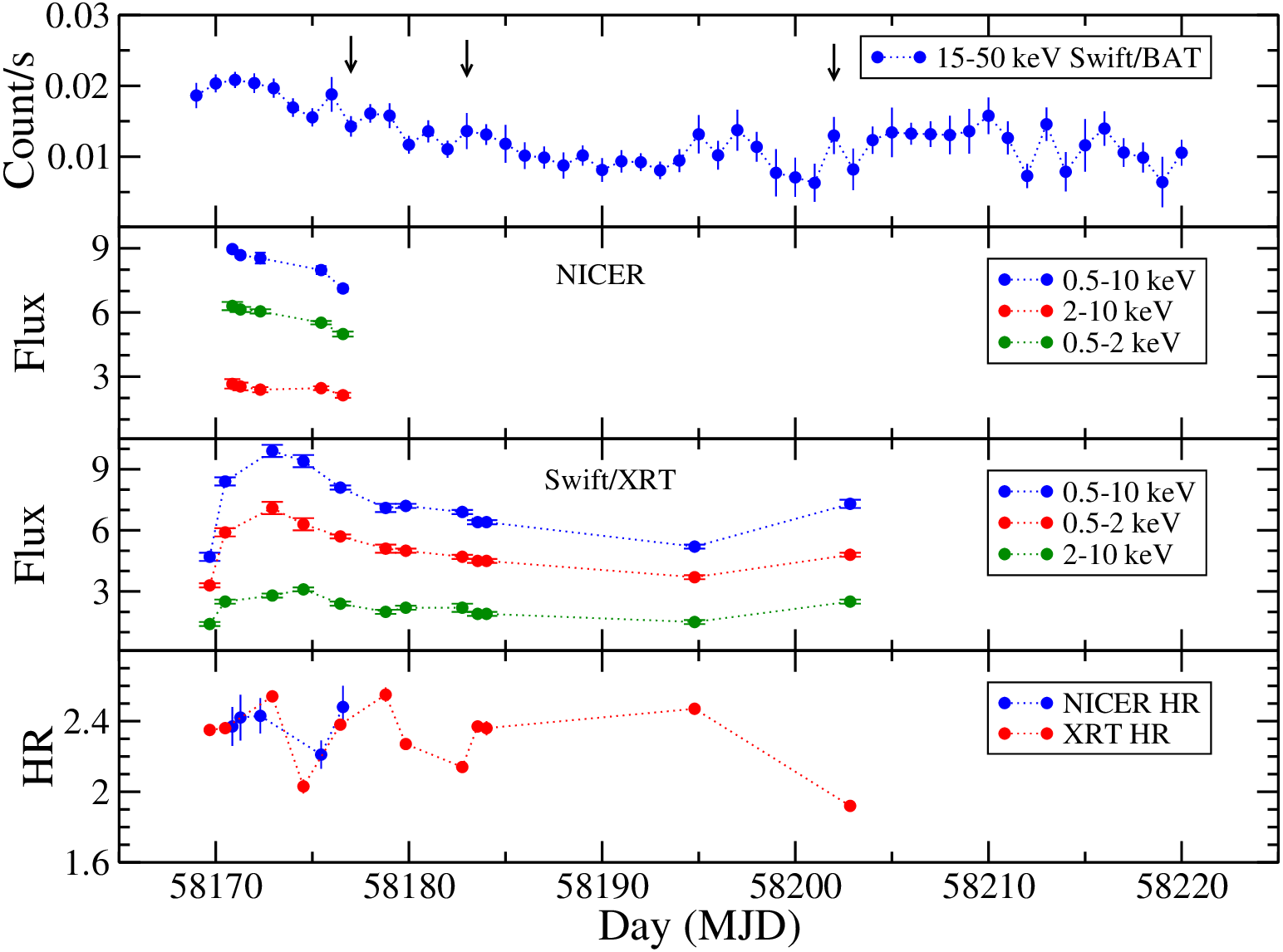}
\caption{The top panel shows the $15-50$~keV {\it Swift}/BAT light curve of MAXI~J1813-095 from 2018 February 16 (MJD 58165) to 2018 April 17 (MJD 58225). The arrows represent the epochs of the {\it NuSTAR} observation of the source. In the second panel, absorption corrected flux in $0.5-10$~keV, $0.5-2$~keV, and $2-10$~keV energy ranges, obtained from {\it NICER} observations are shown. The third panel shows the variation of the $0.5-10$~keV, $0.5-2$~keV, and $2-10$~keV {\it Swift}/XRT unabsorbed flux during the outburst. The fluxes are plotted in the unit of 10$^{-10}$ ergs cm$^{-2}$ s$^{-1}$. In the bottom panel, the hardness ratio (HR) i.e. ratio between fluxes in $2-10$~keV and $0.5-2$~keV ranges, obtained from NICER and XRT data, are shown.}
\label{fig:lc}
\end{figure}

\begin{figure}
\includegraphics[width=10cm]{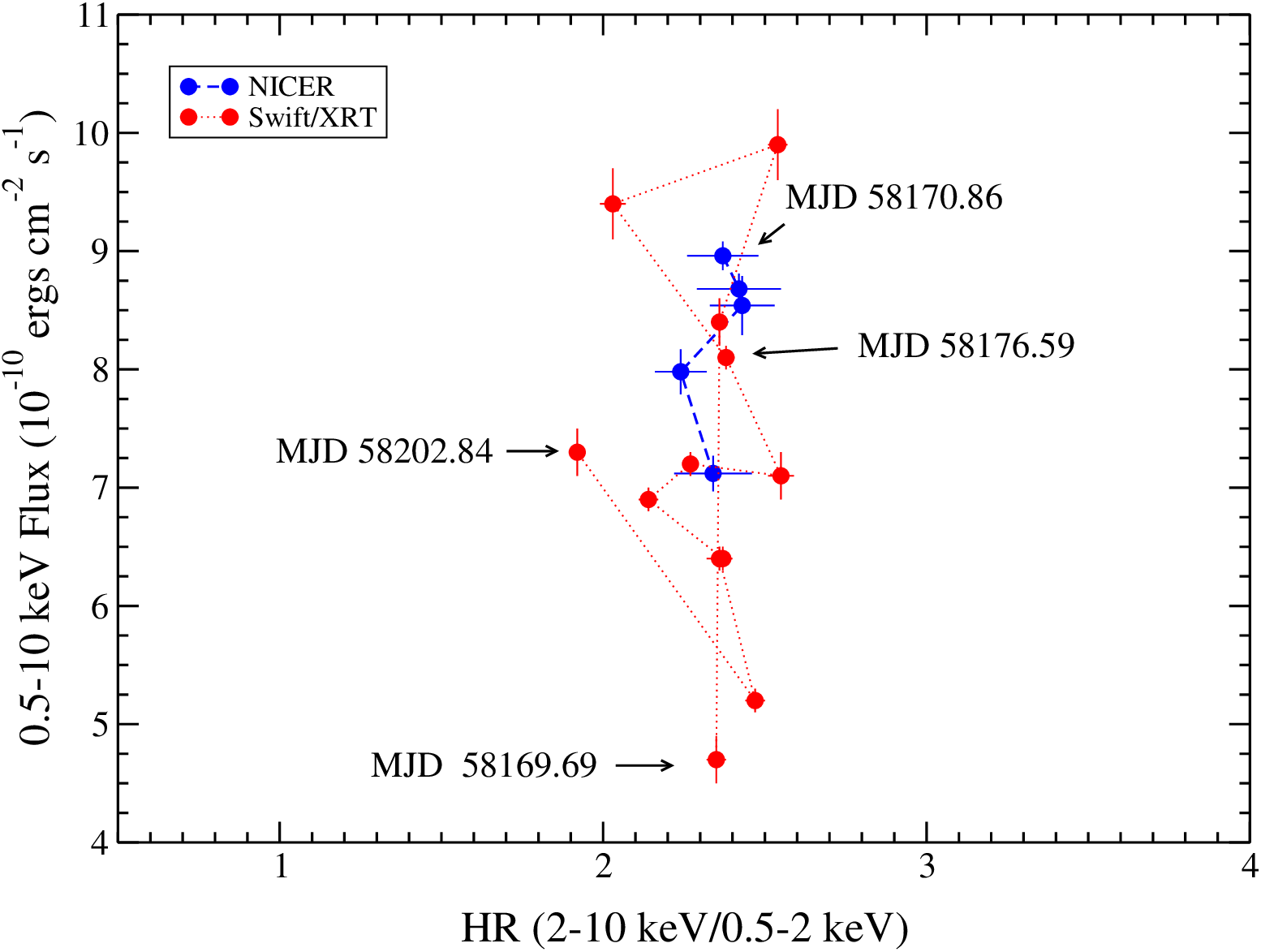}
\caption{Hardness intensity diagram (HID) is shown for $0.5-10$~keV {\it Swift}/XRT and {\it NICER}~ observations. Red and blue points represent {\it Swift}~ and {\it NICER} observations, respectively. Hardness ratio is defined as the ratio between $2-10$~keV flux and $0.5-2$~keV flux. The $0.5-10$~keV flux is in a unit of $10^{-10}$ ergs cm$^{-2}$ s$^{-1}$.}
\label{fig:hid}
\end{figure}

\begin{figure*}
\includegraphics[width=4.8cm]{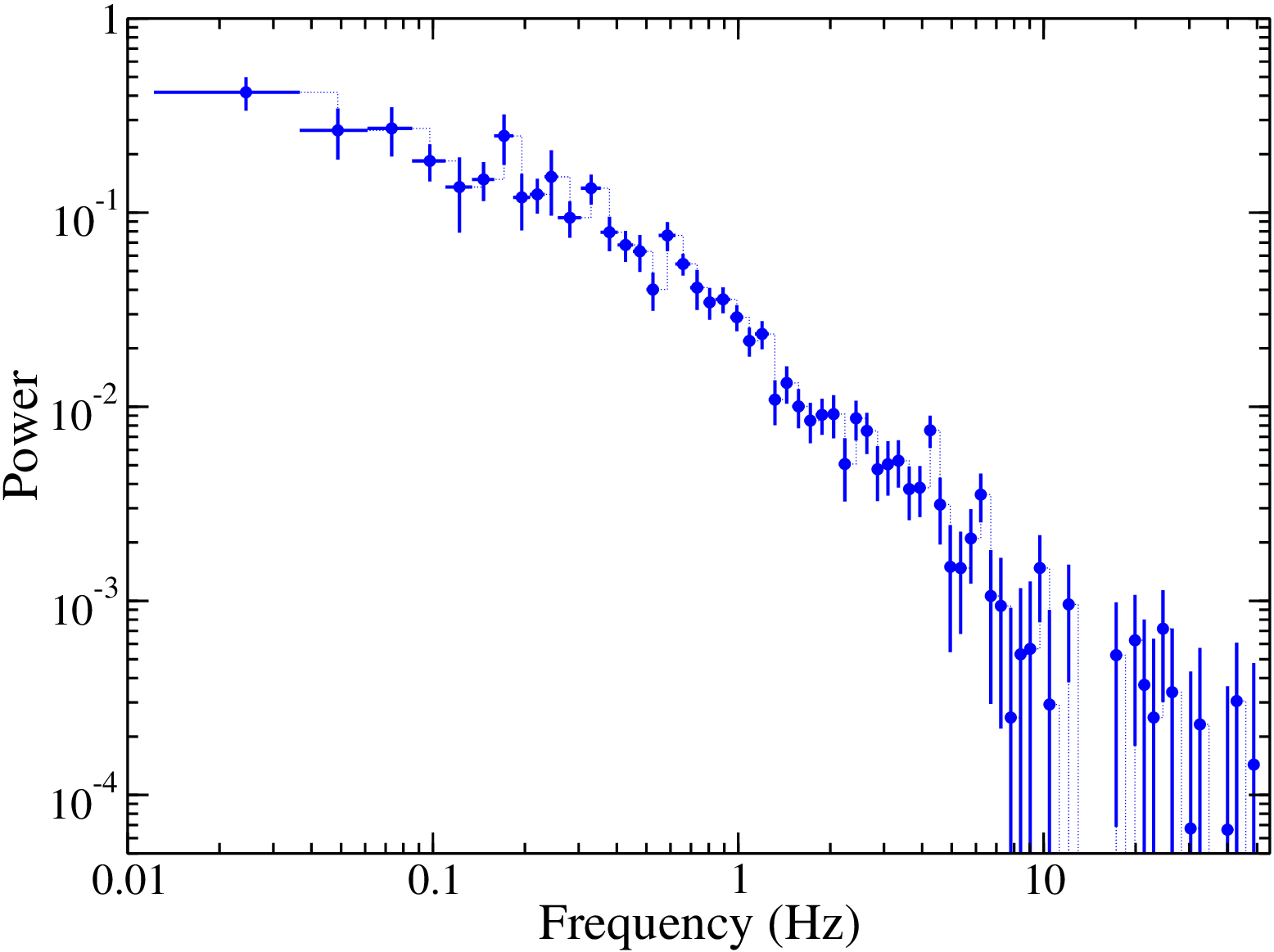}\hskip0.2cm
\includegraphics[width=4.8cm]{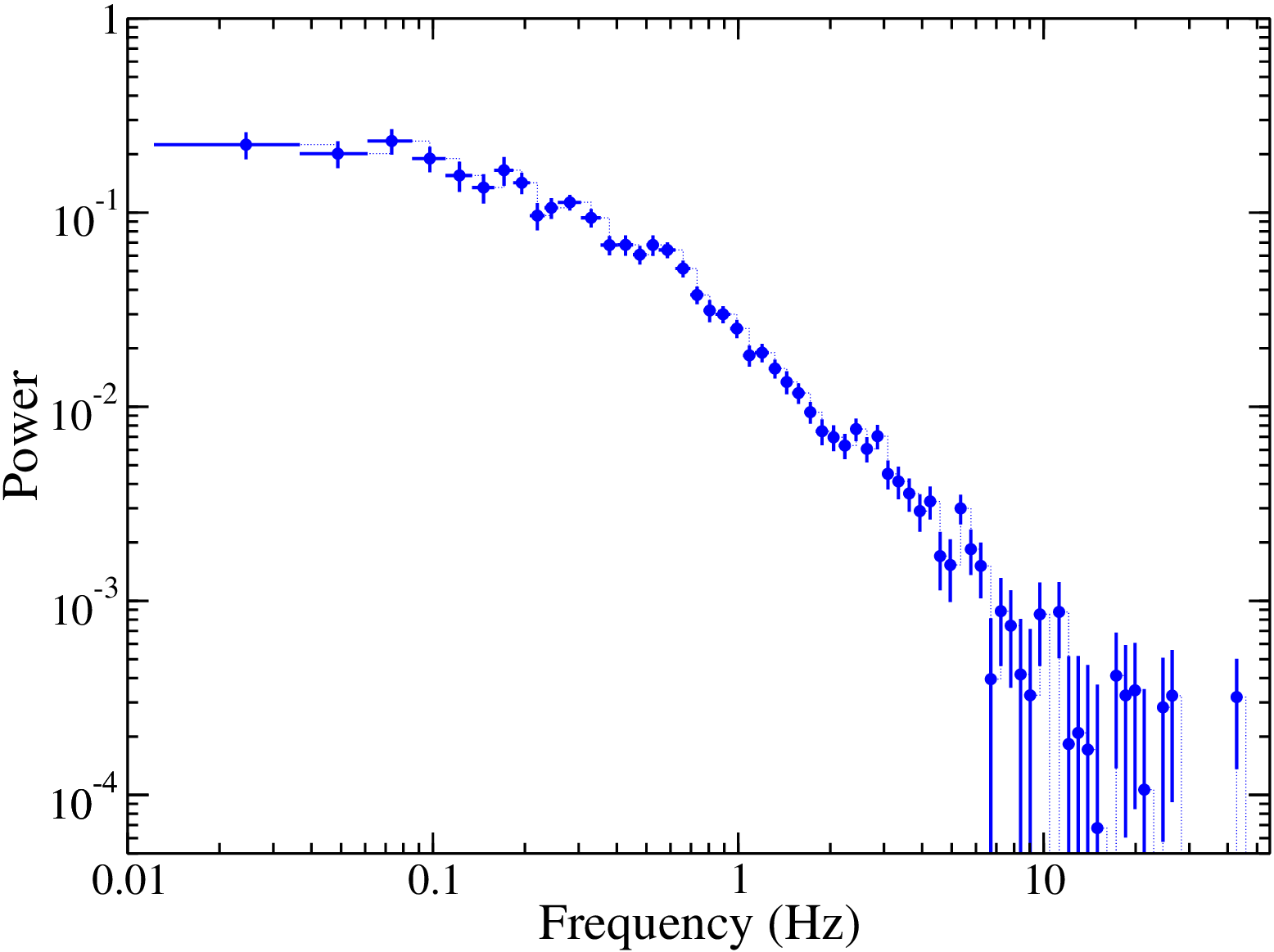}\hskip0.2cm
\includegraphics[width=4.8cm]{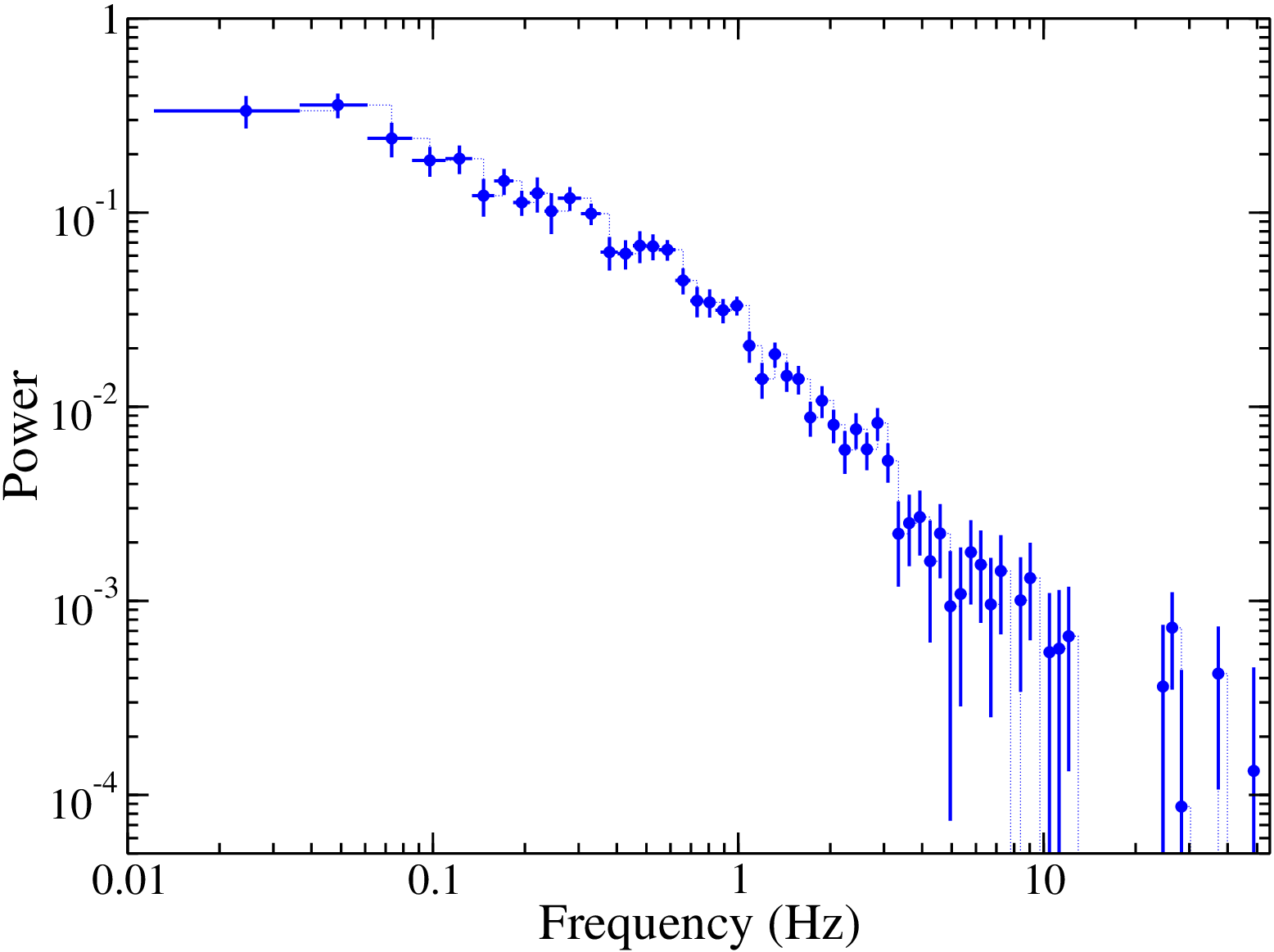}
\caption{The power density spectra (PDS) obtained from the {\it NICER} observations of MAXI~J1813-095
on 2018 February 21 (Obs. ID : 1200090101, left panel), 2018 February 23 (Obs. ID : 1200090103, middle 
panel), and 2018 February 27 (Obs. ID : 1200090105, right panel). The PDS are generated from 0.01 s 
light curves in $0.5-10$~keV energy range.}
\label{fig:pds}
\end{figure*}

\section{Result}
\subsection{Timing Analysis}

The BHC MAXI~J1813-095 was first detected on 2018 February 19 while undergoing the recent X-ray outburst. The outburst lasted for about $\sim 50$ days. The evolution of the outburst is shown in Figure~\ref{fig:lc} using data from {\it Swift}/BAT monitoring, {\it NICER}, {\it Swift}/XRT observations. In the top panel of Figure~\ref{fig:lc}, we show the outburst profile of the source with the {\it Swift}/BAT monitoring light curve in $15-50$~keV energy range. It can be seen that the outburst peaked on 2018 February 22 (MJD 58171) with intensity $\sim$95 mCrab in $15-50$~keV range. The outburst was followed with {\it NICER} from 2018 February 21 (MJD 58170.86), when the source was at its peak with a source count rate of $157 \pm 0.5$ count s$^{-1}$ in $0.5-10$~keV energy range. The second panel of the figure shows the light curve of the source in $0.5-10$, $0.5-2$~keV, and $2-10$~keV energy ranges, obtained from {\it NICER} observations. Third panel of the figure represents absorption corrected source flux in $0.5-2$~keV, $2-10$~keV, and $0.5-10$ keV ranges, estimated from {\it Swift}/XRT data. From Figure~\ref{fig:lc} (second panel), it can be seen that the {\it NICER} observations started when the source was brightest (on 2018 February 21, MJD 58170.86). However, the {\it Swift}/XRT flux, as shown in the third panel, was maximum on 2018 February 23 (MJD 58172.93). It should be noted that the data presented in second and third panels are in the $0.5-10$~keV range. Difference in the outburst peaking times in {\it NICER} and {\it Swift}/XRT data was due to the fact that the source was not observed with the {\it Swift}/XRT between MJD 58170.69 and MJD 58172.93 (between 2018 February 21 and 2018 February 23). The actual peak of the outburst might have been missed in the {\it Swift}/XRT observation. In the bottom panel of Figure~\ref{fig:lc}, the hardness ratio (HR) (ratio between fluxes in $2-10$~keV and $0.5-2$~keV ranges) of the source during the outburst is shown by using NICER and {\it Swift}/XRT data. It can be seen from the figure that the source intensity gradually decreased from 2018 February 21 as the outburst entered the declining phase. A brief re-brightening was observed on 2018 March 25. Soon after that, the source entered to the quiescent state. To investigate the spectral evolution of the source during the outburst, we plotted the hardness-intensity diagram (HID) (source flux vs hardness ratio), obtained from the {\it Swift}/XRT and {\it NICER} observations in $0.5-10$~keV energy range and showed in Fig.~\ref{fig:hid}. The rising (increasing flux) and declining (decreasing flux) phase of the outburst can be traced in the HID (Fig.~\ref{fig:hid}) through the {\it Swift}/XRT data points. It can be seen that the data points in the HID appear to lie in the branch corresponding to the hard state of the ``Q-diagram'' of black hole sources \citep{Homan2005}. The X-ray intensity varied during the outburst, though the HR remained approximately same, indicating no change of spectral states. Considering the evolution of $0.5-2$~keV and $2-10$~keV {\it Swift}/XRT light curves and the HR plot (Figure~\ref{fig:lc}) and the HID (Figure~\ref{fig:hid}), it is clear that the source remained in the hard spectral state during the entire outburst in 2018. 

We analysed the 0.01~s light curves in $0.5-10$~keV range obtained from {\it NICER} observations. White-noise subtracted power density spectrum (PDS) were generated by applying fast Fourier transformation (FFT) technique on the light curves with the {\tt FTOOLS} task {\tt powspec norm=-2} for different intervals such as 2048, 4096 and 8192.  In Figure~\ref{fig:pds}, we show the PDSs generated from the 0.01~s light curves from the {\it NICER} observations on (a) 2018 February 21 (Obs ID: 1200090101), (b) 2018 February 23 (Obs ID: 1200090103) and (c) 2018 February 27 (1200090105). The 0.01~s binning time allowed us to search for presence/absence of QPOs up to 50 Hz in each PDS. However, we did not find any signature of presence of QPO in the PDS of any of the observations. All the PDSs showed weak red-noise with flat top noise up to $0.1$ Hz. A strong rms is observed in all the PDSs with rms$ \simeq 20-30$ \% in $0.1-50$~Hz band. We also investigated the presence of QPOs in the PDS obtained from the {\it Swift}/XRT light curves, and obtained similar results. We attempted to search for the signature of high frequency QPOs by generating PDS from light curves with 0.004 s time bin from {\it NICER} observations. However, as in case of search for low frequency QPOs, there was no signature of presence of any high frequency QPOs in the PDS up to 1250 Hz.

\subsection{Spectral Analysis}

We study the BHC MAXI~J1813-095 during its 2018 outburst using data from {\it Swift}/XRT, {\it NICER}, and {\it NuSTAR} observations in the energy range of $0.5-78$~keV. We carry out spectral analysis with HEASARC's spectral analysis software package {\tt XSPEC} v12.10\footnote{\url{https://heasarc.gsfc.nasa.gov/docs/xanadu/xspec/}} \citep{Arnaud1996}. For interstellar absorption, we used {\tt TBabs} model with Wilms abundances \citep{Wilms2000}.

\subsubsection{{\it Swift}}

MAXI~J1813-095 was observed with the {\it Swift} observatory at twelve epochs during the 2018 X-ray outburst. The source and background spectra, effective area and response files were generated as described in the previous section, and used in the spectral fitting. The $0.5-10$~keV XRT spectra were fitted well with an absorbed power law model. We fixed the hydrogen column density ($N_H$) at $1.1\times 10^{22}$ cm$^{-2}$ \citep{AP19}. The power-law photon index ($\Gamma$) was found to vary between 1.54 and 1.68 during the outburst period. We also calculated the unabsorbed flux in $0.5-2$~keV and $2-10$~keV energy bands using `{\tt cflux}' command in {\tt XSPEC}. In Figure~\ref{fig:pl}a (left panel), we show a representative $0.5-10$~keV {{\it Swift}}/XRT spectrum fitted with a {\tt powerlaw} model, observed on 2018 February 25 (Obs ID : 00010563004).

\subsubsection{{\it NICER}}

\begin{figure*}
\includegraphics[width=4.8cm]{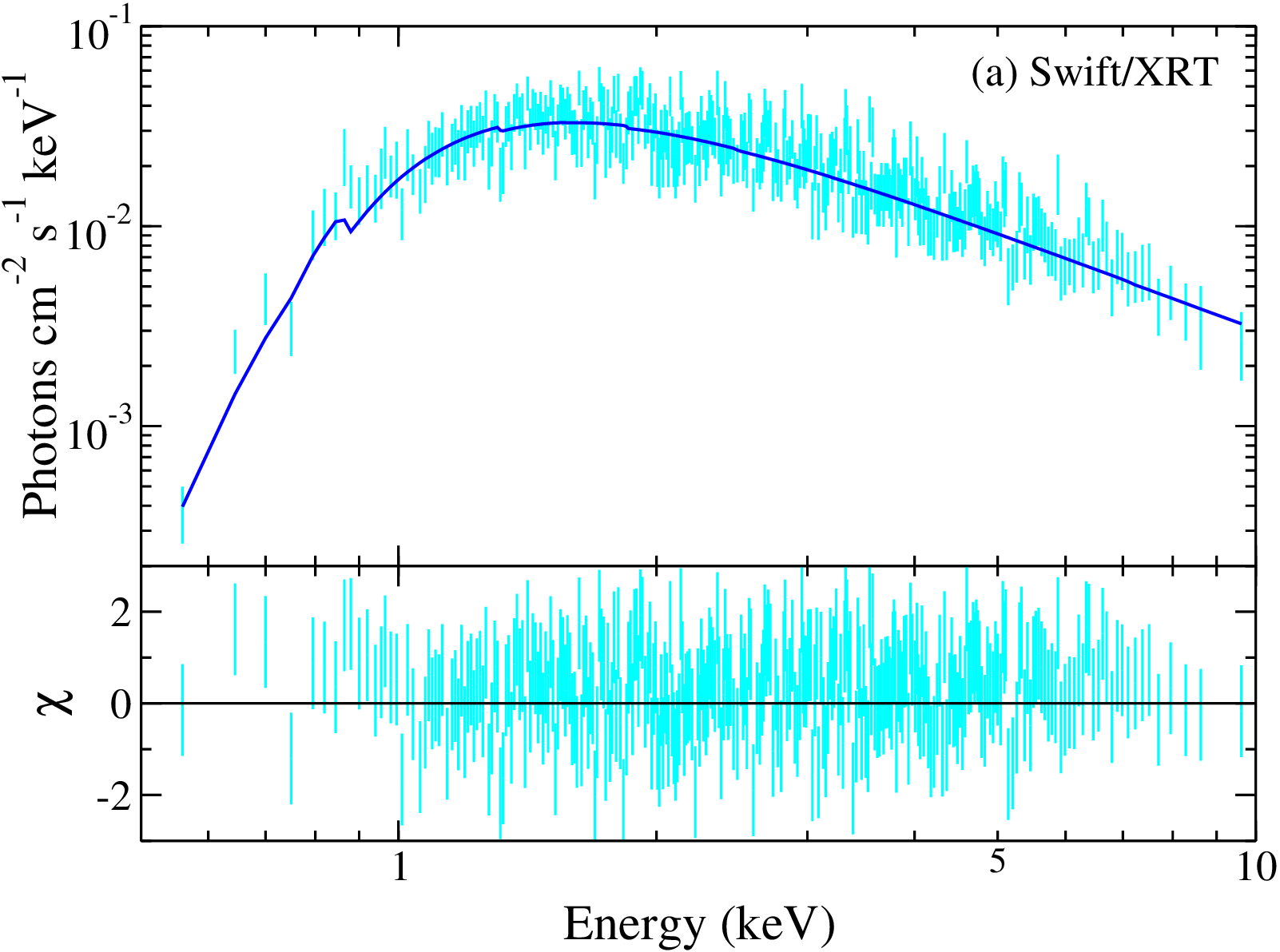}
\includegraphics[width=4.8cm]{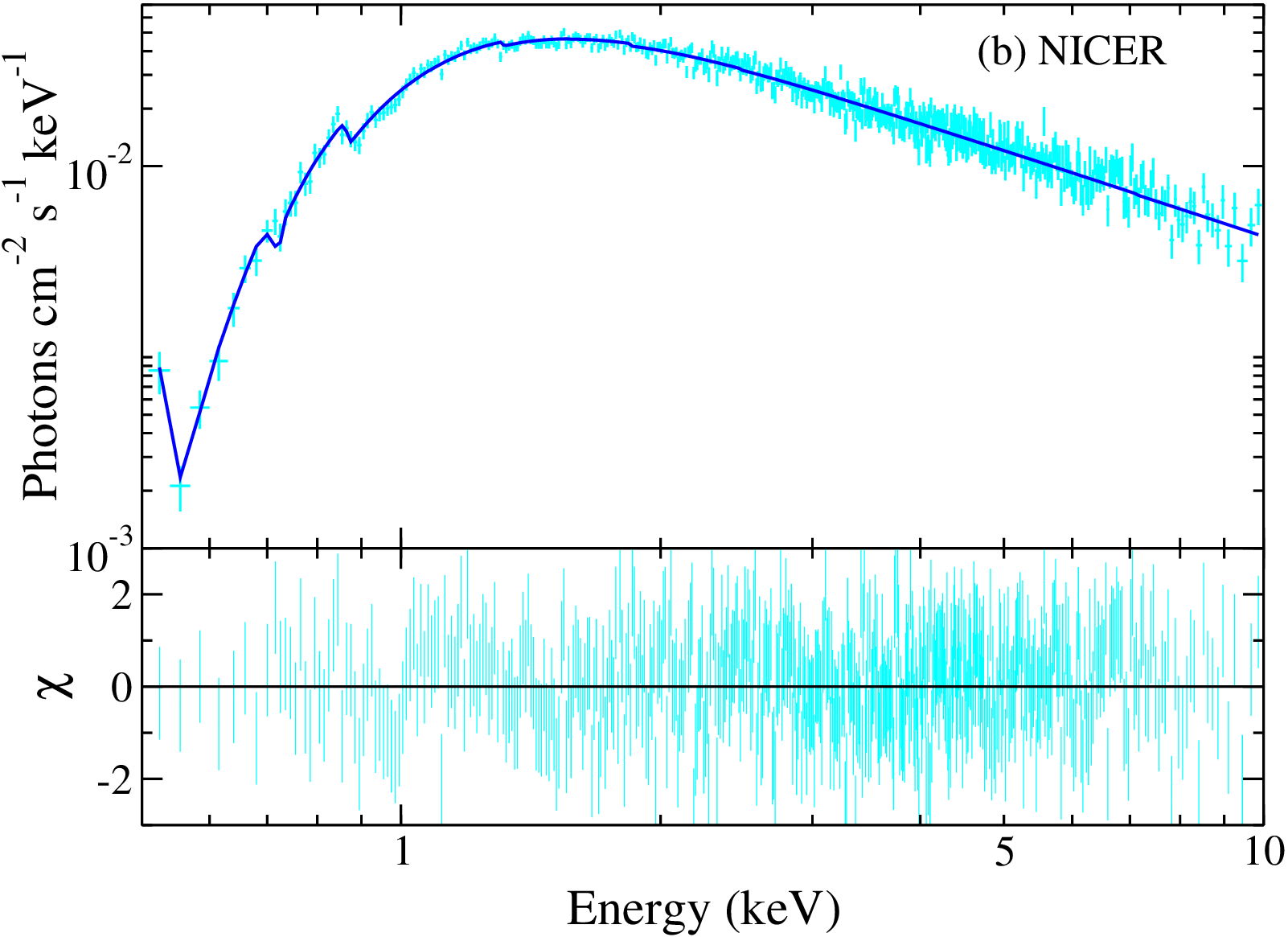}
\includegraphics[width=4.8cm]{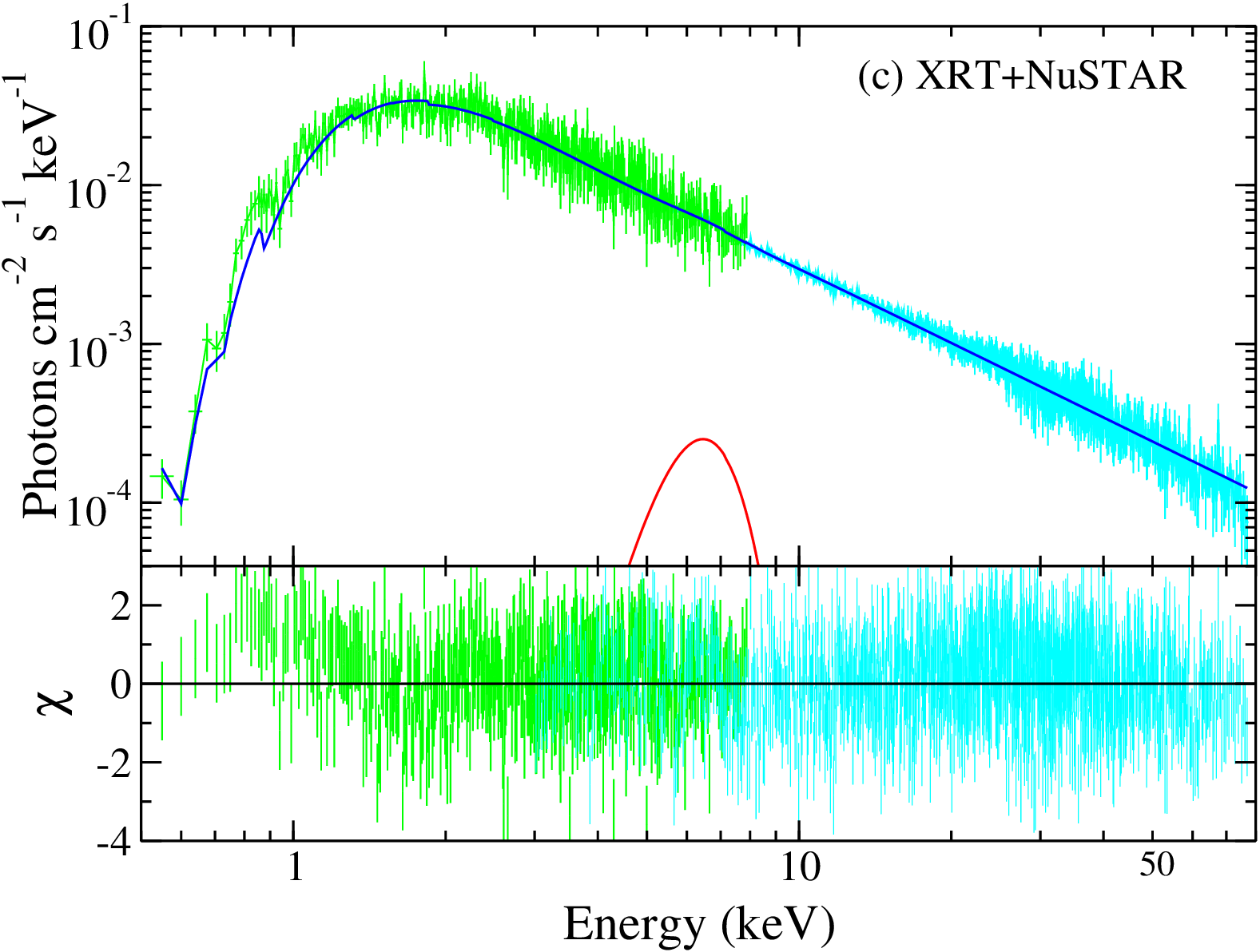}
\caption{Spectra of the BHC MAXI~J1813-095 along with the best-fit model and residuals obtained 
from the (a) {\it Swift}/XRT observation on 2018 February 25 (Obs ID: 00010563004) in $0.5-10$~keV 
range (left panel), (b) {\it NICER} observation on 2018 February 22 (Obs ID: 1200090102) in $0.5-10$~keV 
range (middle panel), and (c) {\it Swift}/XRT and {\it NuSTAR} simultaneous observations on 2018 
March 6 ({\it Swift} Obs ID: 00088654002; {\it NuSTAR} Obs ID: 80402303004) in $0.5-78$~keV range 
(right panel), are shown. The $0.5-10$~keV {\it Swift}/XRT (left panel), $0.5-10$~keV {\it NICER}
(middle panel), and $0.5-78$~keV {\it Swift}/XRT+{\it NuSTAR} (right panel) spectra were fitted 
with {\tt TBabs*powerlaw}, {\tt TBabs*(diskbb+powerlaw}), and {\tt TBabs(diskbb+powerlaw+Gaussian)} 
model, respectively. The residuals obtained from the spectral fitting are shown in the bottom panels.}
\label{fig:pl}
\end{figure*}

{\it NICER} observed MAXI~J1813-095 five times during the 2018 outburst. The $0.5-10.0$~keV spectra were fitted with the absorbed power law ({\tt powerlaw}) model along with the disk-blackbody ({\tt diskbb}) component. Spectra from all the observations were well fitted with this model. The inner disk temperature ($T_{in}$) varied between $0.54 - 0.61$~keV along with a approximately constant power-law photon index ($\Gamma$) ($\sim 1.52-1.55$). No signature of Fe K$\alpha$ line was observed in the {\it NICER} data. We show a representative $0.5-10$~keV {\it NICER} spectrum in Figure~\ref{fig:pl}b (middle panel), observed on 2019 February 22 (Obs ID: 1200090102). The {\tt powerlaw + diskbb} model fitted spectral parameters are shown in Table~\ref{tab:nicer}.

\begin{table}
\caption{Spectral fitting parameters obtained from  the {\it NICER} Observations}
\label{tab:nicer}
\begin{tabular}{lccccccc}
\hline
ID	 &$T_{in}$ & DBB & $\Gamma$ & Flux & $\chi^2$/dof \\
 & (keV) &Norm. &  & & \\
\hline
X1 &0.57$\pm$0.06  &43.8$\pm$2.1 &1.53$\pm$0.05  &8.96$\pm$0.12  &552/580 \\
X2 &0.56$\pm$0.05  &56.1$\pm$2.4 &1.52$\pm$0.06  &8.68$\pm$0.11  &569/537 \\
X3 &0.54$\pm$0.03  &40.5$\pm$2.0 &1.54$\pm$0.03  &8.54$\pm$0.15  &797/750 \\
X4 &0.58$\pm$0.09  &16.1$\pm$1.2 &1.55$\pm$0.04  &8.22$\pm$0.19  &592/628 \\
X5 &0.61$\pm$0.05  &18.8$\pm$1.2 &1.52$\pm$0.04  &8.12$\pm$0.15  &672/665 \\
\hline
\end{tabular}

\leftline{$N_H$ was fixed at 1.1$\times$ 10$^{-22}$ cm$^{-2}$. Errors are quoted with 90\% confidence.}
\leftline{Flux is in unit of $10^{-10}$ ergs~cm$^{-2}$~s$^{-1}$ and estimated in $0.5-10$~keV energy range.}
\end{table}

\subsubsection{Swift + NuSTAR}

{\it NuSTAR} observed MAXI~J1813-095 three times during the 2018 X-ray outburst. Among those, two observations were made simultaneously with the {\it Swift}/XRT. We attempted to carry out simultaneous spectral fitting of {\it Swift}/XRT and {\it NuSTAR} data with an absorbed power law model. However, fitting the broad-band spectra with the absorbed power law model did not provide us satisfactory fitting with $\chi^2 = 1470$ for 1122 dof for the {\it NuSTAR} observation on 2018 February 28 (N1 in Table~1). Signatures of a disk and Fe K$\alpha$ emission line were seen in the residuals. Adding a {\tt diskbb} component with the model improves the fit with $\chi^2 = 1328$ for 1120 dof. We further added a Gaussian function for the Fe K$\alpha$ line, which significantly improves the fit with $\chi^2 = 1155$ for 1117 dof. The other two {\it NuSTAR} observations (N2 \& N3 in Table~\ref{tab:log}) when fitted along with simultaneous {\it Swift}/XRT data (S1 \& S2 in Table~\ref{tab:log}), also showed similar results. Therefore, the {\tt TBabs*(diskbb+powerlaw+Gaussian)} model fits well the broad-band spectra of MAXI~J1813-095 from {\it NuSTAR} and {\it Swift}/XRT observations. The power-law photon index ($\Gamma$) was found to be 1.56, 1.57 and 1.62 for N1, N2+S1 and N3+S2, respectively. The inner disk temperature ($T_{in}$) varied between 0.61~keV and 0.57~keV. It was found that during all three {\it NuSTAR} observations, the power-law flux dominated over the thermal flux. The fraction of the thermal flux was less than $\sim$10\% of total flux in $0.5-78$~keV range, obtained from simultaneous fitting of {\it Swift}/XRT and {\it NuSTAR} data, and less than 1\% of total flux in $3-78$~keV energy range obtained from fitting of {\it NuSTAR} data (N1). The best-fit model parameters are given in Table~\ref{tab:res}.

We often observed a reflection hump at around $\sim 15-30$~keV in the hard state spectra \citep{GF1991,Matt1991}. Often, the presence of reflection makes the spectra harder. Unusual hard spectra are observed in MAXI~J1813-095 with low photon index. In order to probe the spectral nature and reflection continuum further, we explore the `reflection' with convolution model for reflection {\tt reflect} \citep{MZ95}. This model describes the reflection from relatively cold neutral material. We fixed heavy element abundances and iron abundances at Solar value (i.e. 1). We allowed the relative reflection ($R_{refl}$), photon index ($\Gamma$) and inclination angle of the system (as cos incl) to vary. All three observations yield marginally improved fit compared to the {\tt powerlaw} model fitting. The photon index $\Gamma$ was constant at around 1.65. The $R_{refl}$ was 0.15, 0.22 and 0.25 for N1, N2+S1, and N3+S2, respectively. The $Cos(incl)$ varied between 0.71 and 0.88, which transformed the inclination angle between $28^{\circ}$ and $45^{\circ}$. The inner disk temperature ($T_{in}$) was observed to be 0.56, 0.48, and 0.40 keV for N1, N2+S1, and N3+S2, respectively.

\begin{figure*}
\includegraphics[width=4.8cm]{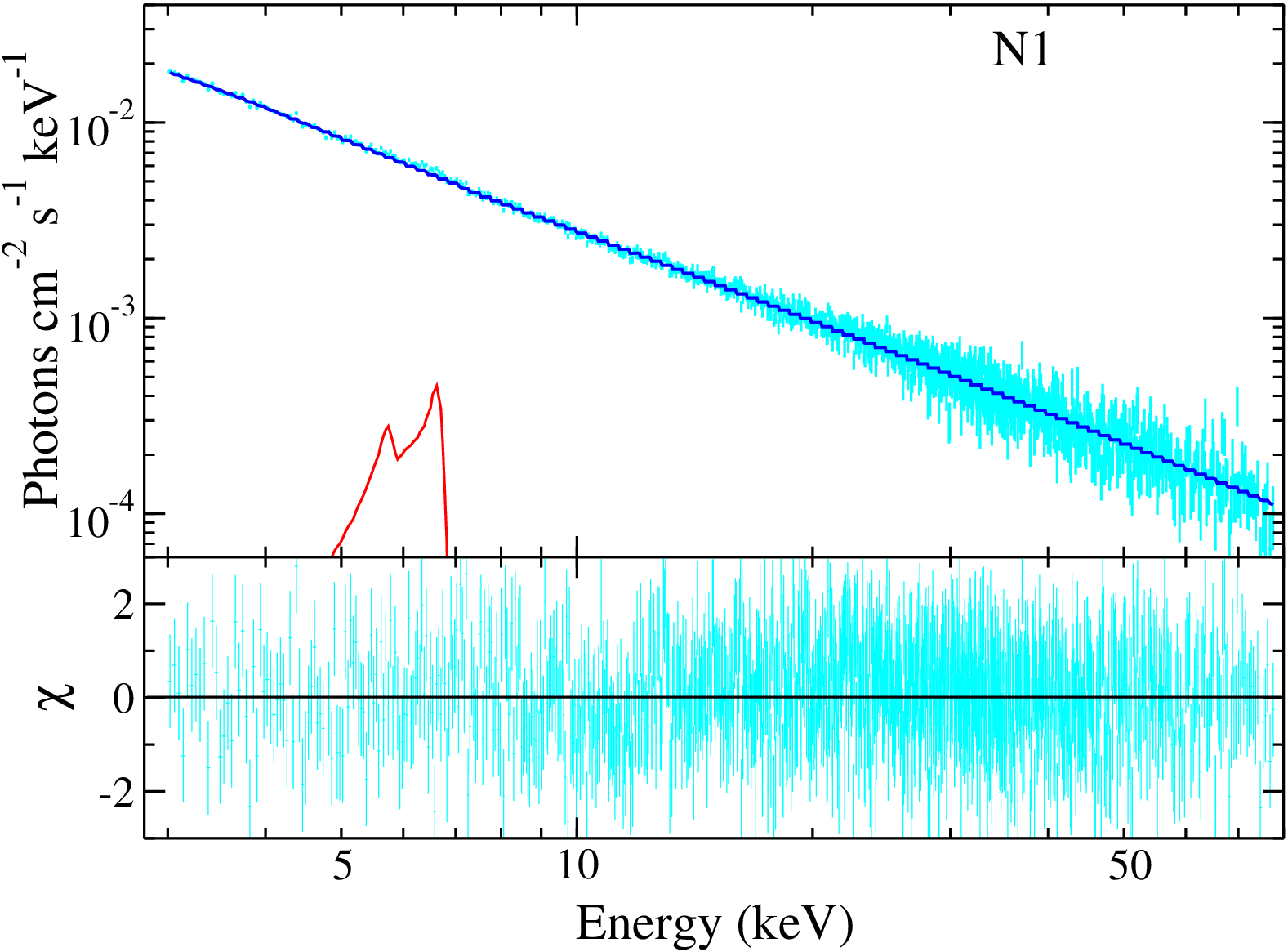}
\includegraphics[width=4.8cm]{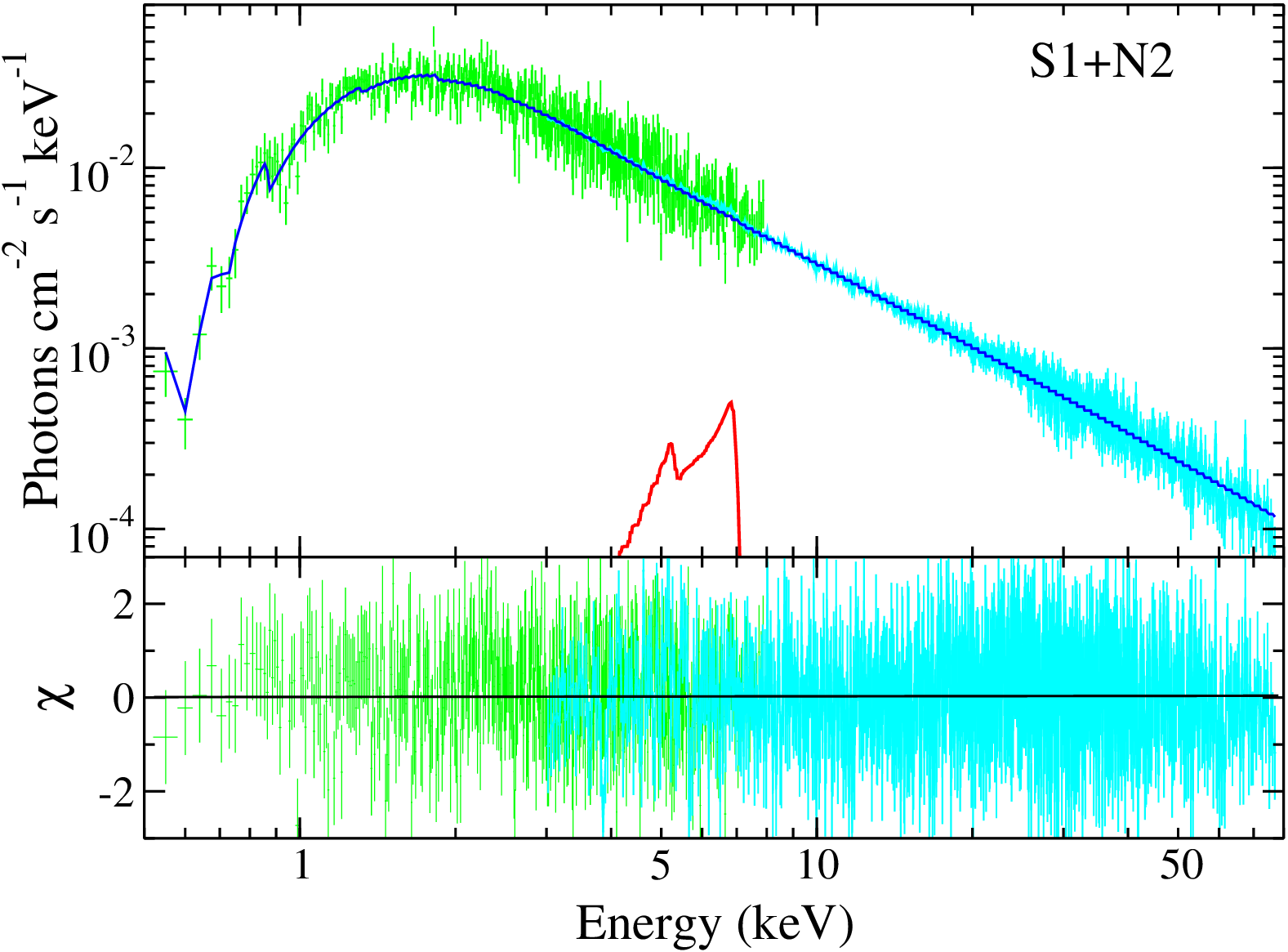}
\includegraphics[width=4.8cm]{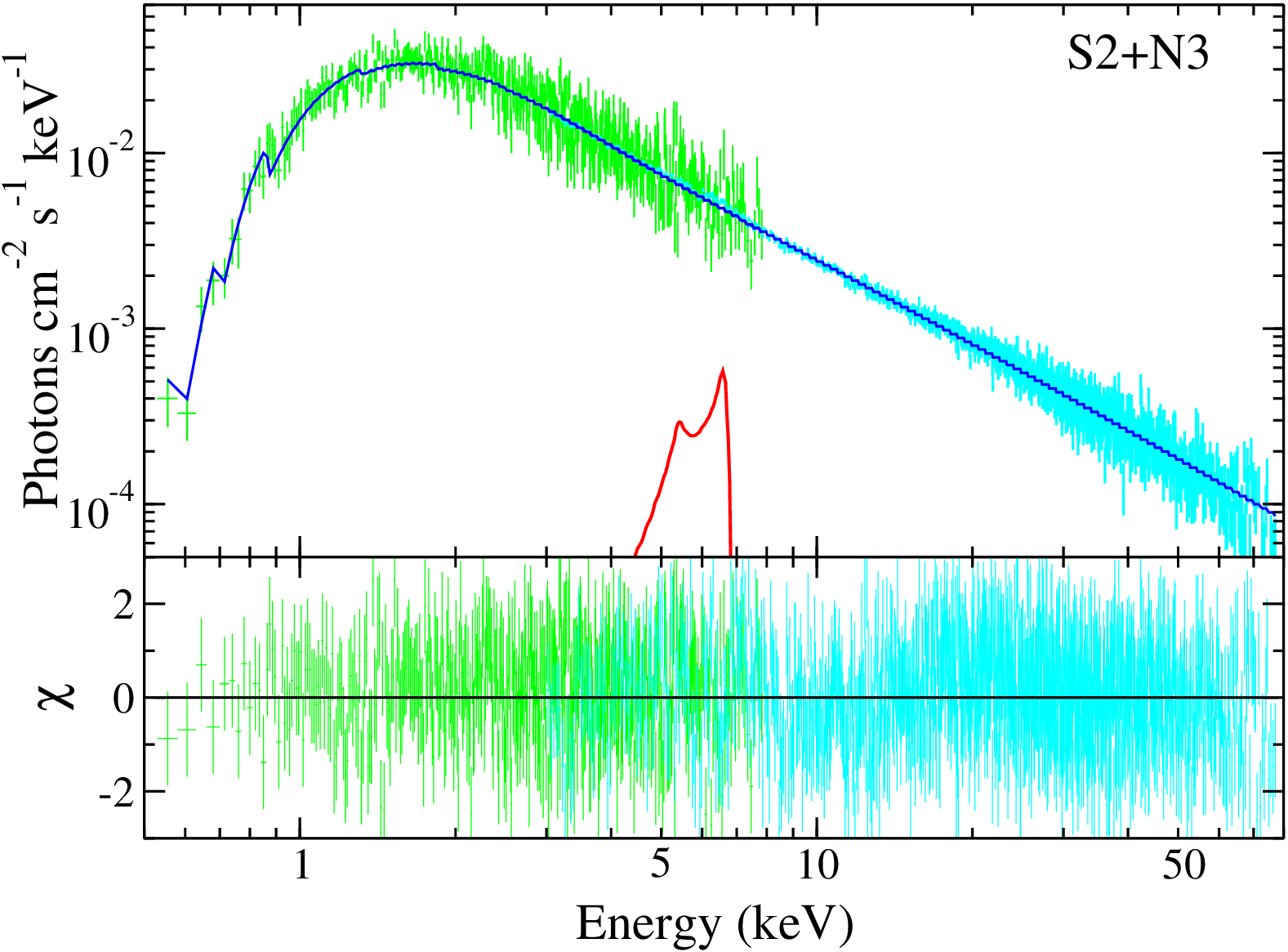}
\caption{The $3-78$~keV (left panel) and $0.5-78$~keV (middle and right panels) TCAF+LAOR model fitted spectra and residuals are shown for first {\it NuSTAR} (N1), {\it NuSTAR}+{\it Swift}/XRT (N2+S1 and N3+S2) observations, respectively.}
\label{fig:spec}
\end{figure*}

\begin{figure}
\centering
\includegraphics[width=10cm]{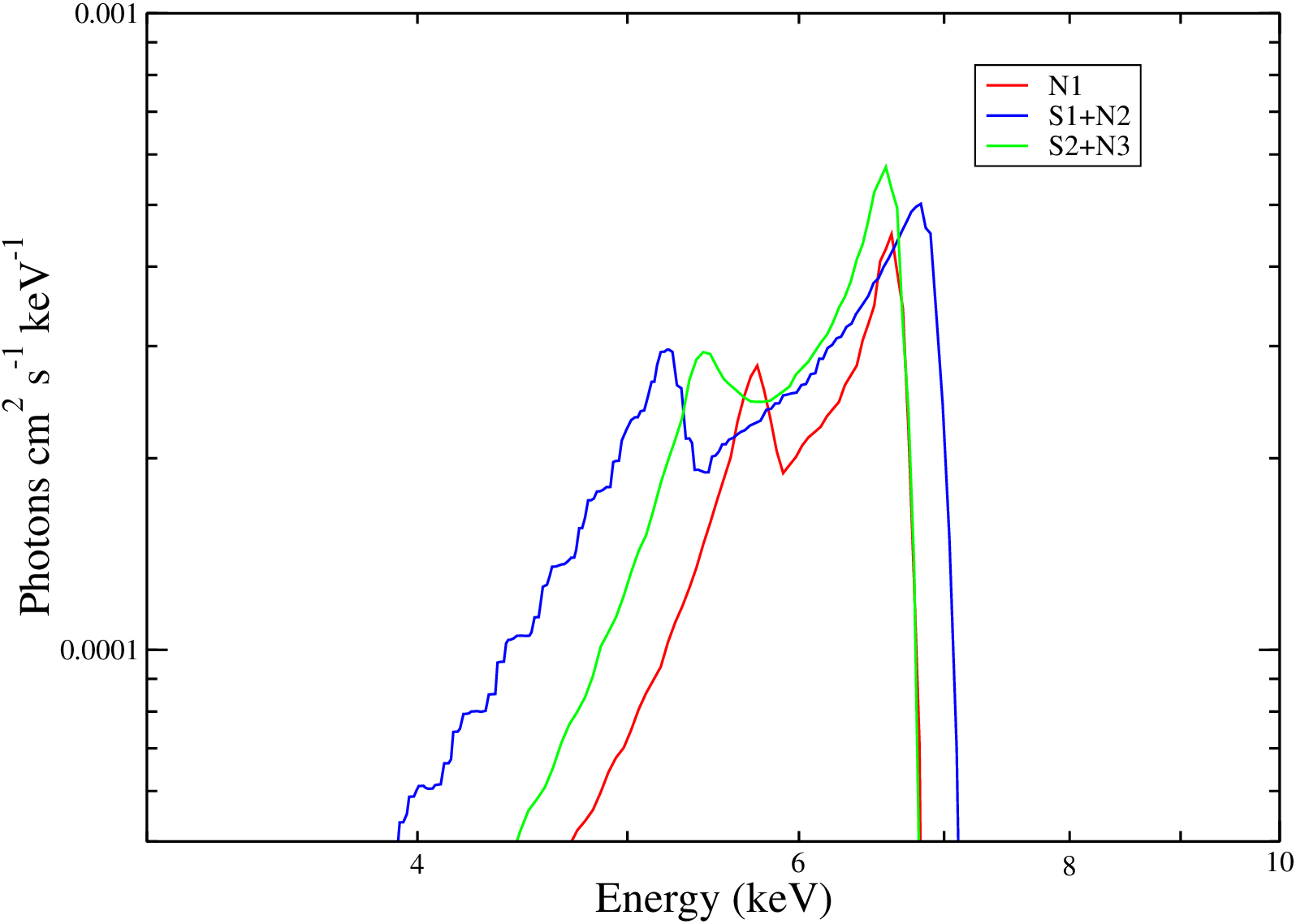}
\caption{The Fe K$\alpha$ line intensity is shown for three {\it NuSTAR} observations.}
\label{fig:laor}
\end{figure}

Next, we used physical model TCAF as a local additive model in {\tt XSPEC} \citep{DD14,DD15}. Along with the TCAF, we used {\tt LAOR} model \citep{Laor91} to incorporate the iron K$\alpha$ emission line. The TCAF model has five input parameters: the mass of the black hole ($M_{BH}$) in Solar mass ($M_{\sun}$), the Keplerian disk accretion rate ($\dot{m}_d$) in Eddington rate ($\dot{M}_{Edd}$), the sub-Keplerian halo accretion rate ($\dot{m}_h$) in Eddington rate ($\dot{M}_{Edd}$), the shock location or the size of the Compton cloud ($X_s$) in Schwarzschild radius ($r_s$), and the shock compression ratio ($R$, ratio between post-shock matter density to pre-shock matter density). Along with these, we obtain normalization ($N$), which is a function of mass of the BH, the distance of the source and inclination angle of the system. As these three parameters are intrinsic to the system, the normalization parameter $N$ should remain unchanged during the outburst \citep{AJ17,AJ20a}. The {\tt TCAF+LAOR} model gave us a good fit for all three observations. The TCAF model fitted spectra are shown in Figure~\ref{fig:spec}. We show the Fe K$\alpha$ line intensity in Figure~\ref{fig:laor}. 

While fitting the data with the TCAF model, we kept mass of the BH as a free parameter. We obtained $M_{BH}$ as 7.38, 7.44, and 7.40 $M_{\sun}$from N1, N2+S1 and N3+S2, respectively. The Keplerian disk mass accretion rate $\dot{m}_d$ varied between 0.07 $\dot{M}_{Edd}$ and 0.05 $\dot{M}_{Edd}$. The sub-Keplerian halo accretion rate $\dot{m}_h$ varied between 0.54 $\dot{M}_{Edd}$ and 0.52 $\dot{M}_{Edd}$. The dominance of the sub-Keplerian flow indicates the hard spectral state of the source during the observation period. We also observed that the shock moved outward from 93 $r_s$ to 113 $r_s$. Thus, the size of the Compton cloud increased as the outburst progressed. The shock was strong during all three observations with the shock compression ratio $R \sim 2.80$. During all three observations, the normalization was roughly constant with $N \sim 1.65$. The TCAF model fitted results are shown in Table~\ref{tab:res}. We used {\tt LAOR} model along with the TCAF for a relativistic broad iron line. The broad Fe K$\alpha$ line is shown in Figure~\ref{fig:laor}.

\begin{table}
\caption{Best-fit spectral parameters obtained from {\it NuSTAR} and {\it Swift}/XRT observations}
\label{tab:res}
\begin{tabular}{lccccc}
\hline
Model       &Parameter  &{\it NuSTAR}  &\multicolumn{2}{c}{{\it NuSTAR+Swift}/XRT}\\
Comp.	    &           & N1           & N2+S1   & N3+S2 \\
\hline
	diskbb&$T_{in}$ (keV)& $0.62^{+0.07}_{-0.05}$ & $0.61^{+0.04}_{-0.10}$ & $0.57^{+0.05}_{-0.08}$\\
	&norm& $92.6^{+8.4}_{-7.9}$& $103.6^{+12.9}_{-14.2}$ & $121.0^{+15.5}_{-12.8}$ \\
	Powerlaw& $\Gamma$& $1.56^{+0.05}_{-0.04}$ & $1.57^{+0.05}_{-0.05}$& $1.62^{+0.04}_{-0.08}$ \\
	& norm & $0.15^{+0.03}_{-0.03}$& $0.11^{+0.02}_{-0.03}$ & $0.10^{+0.01}_{-0.01}$\\
	Gaussian&LE (keV)& $6.20^{0.25}_{-0.22}$& $6.47^{+0.27}_{-0.24}$ & $6.23^{+0.19}_{-0.28}$\\
	&$\sigma$ (keV)& $0.77^{+0.06}_{-0.09}$ & $0.97^{+0.10}_{-0.07}$ & $0.86^{+0.06}_{-0.09}$\\
	&norm$^*$ & $9.47^{+0.24}_{-0.29}$& $3.77^{+0.29}_{-0.22}$ & $7.21^{+0.33}_{-0.21}$ \\
	&$\chi^2$/dof& 1155/1117& 1610/1555& 1699/1558\\
\hline
	diskbb & $T_{in}$ (keV)& $0.56^{+0.05}_{-0.08}$& $0.48^{+0.04}_{-0.06}$& $0.40^{+0.06}_{-0.10}$ \\
	& norm & $66.5^{+2.9}_{-4.4}$ & $138.7^{+5.5}_{-7.9}$ & $145.6^{+6.5}_{-8.8}$ \\
	Powerlaw& $\Gamma$ & $1.64^{+0.04}_{-0.08}$ & $1.66^{+0.03}_{-0.07}$ & $1.65^{+0.05}_{-0.06}$ \\
	& norm & $0.12^{+0.01}_{-0.01}$ & $0.13^{+0.02}_{-0.03}$ & $0.13^{+0.02}_{-0.02}$ \\
	Reflect &$R_{refl}$& $0.15^{+0.02}_{-0.03}$ & $0.22^{+0.03}_{-0.04}$ & $0.25^{+0.02}_{-0.04}$ \\
	& $\cos$(incl) & $0.88^{+0.02}_{-0.04}$& $0.71^{+0.05}_{-0.06}$& $0.85^{+0.03}_{-0.07}$ \\
	Gaussian & LE (keV) & $6.20^{+0.18}_{-0.22}$ &$6.56^{+0.07}_{-0.20}$ & $6.33^{+0.016}_{-0.23}$ \\
	& $\sigma$ (keV)&$0.78^{+0.20}_{-0.13}$ &$0.97^{+0.12}_{-0.14}$ & $0.81^{+0.12}_{-0.15}$ \\
	& norm$^*$ & $2.51^{+0.25}_{-0.17}$ & $2.06^{+0.26}_{-0.28}$ & $1.28^{+0.20}_{-0.24}0$ \\
	& $\chi^2$/dof& 1126/1115&1555/1553 & 1589/1556 \\
\hline
	TCAF & $M_{BH}$ ($M_{\sun}$)& $7.38^{+1.43}_{-1.46}$ & $7.44^{+1.24}_{-1.56}$ & $7.40^{+1.33}_{-1.33}$ \\
	& $\dot{m}_d$ ($\dot{M}_{Edd}$)&$0.07^{+0.01}_{-0.01}$ & $0.05^{+0.01}_{-0.01}$ & $0.06^{+0.01}_{-0.02}$ \\
	& $\dot{m}_h$ ($\dot{M}_{Edd}$)&$0.54^{+0.03}_{-0.04}$ & $0.51^{+0.02}_{-0.05}$ & $0.52^{+0.04}_{-0.06}$ \\
	& $X_s$ ($r_s$)&$93^{+9}_{-11}$ &$111^{+10}_{-14}$ & $113^{+9}_{-11}$\\
	& $R$ & $2.80^{+0.15}_{-0.16}$ & $2.82^{+0.10}_{-0.18}$& $2.79^{+0.12}_{-0.16}$\\
	& $N_{tcaf}$& $1.62^{+0.08}_{-0.11}$ & $1.65^{+0.07}_{-0.12}$& $1.67^{+0.09}_{-0.12}$ \\
	LAOR & LE (keV)& $6.45^{+0.22}_{-0.16}$ &$6.59^{+0.16}_{-0.23}$ &$6.43^{+0.16}_{-0.20}$ \\
	& index & $1.75^{+0.03}_{-0.04}$&$1.89^{+0.06}_{-0.07}$ &$1.87^{+0.06}_{-0.09}$ \\
	& $R_{in}$ ($r_g$)& $2.64^{+0.10}_{-0.06}$ &$2.59^{+0.11}_{-0.08}$ &$2.58^{+0.11}_{-0.08}$ \\
	& $R_{out}$ ($r_g$)& $68.1^{+3.1}_{-5.5}$& $60.2^{+5.2}_{-6.5}$& $75.6^{+2.5}_{-3.6}$ \\
	& $\theta_{incl}$ (deg)& $35.08^{+1.42}_{-2.77}$&$36.26^{+1.48}_{-2.65}$ & $31.88^{+1.65}_{-2.23}$\\
	& norm$^*$ & $5.33^{+0.41}_{-0.37}$& $8.02^{+0.38}_{-0.45}$& $7.07^{+0.32}_{-0.46}$\\
	& $\chi^2$/dof&1129/1110 &1583/1555 & 1595/1556\\
\hline
    $F_{0.1-100}^{a}$ & $0.1-100$~keV& $2.44^{+0.15}_{-0.11}$ & $1.86^{+0.09}_{-0.12}$ & $1.94^{+0.12}_{-0.10}$ \\
\hline
	$f_{th}^{b}$& $0.5-78$~keV & --- & 5.4\%& 10\%\\
	&$3-78$~keV &0.8\% &0.7\% & 0.9\%\\
\hline
\end{tabular}

\leftline{$^*$ : in the unit of $10^{-4}$ ph~cm$^{-2}$~s$^{-1}$.}
\leftline{$^a$ : in the unit of $10^{-8}$ ergs s$^{-1}$ cm$^{-2}$.}
\leftline{$^{b}$ Thermal fraction, defined as $F_d /(F_d + F_{PL})$. All errors are quoted with 90\% confidence level.}
\end{table}

\begin{figure}
\centering
\includegraphics[width=10cm]{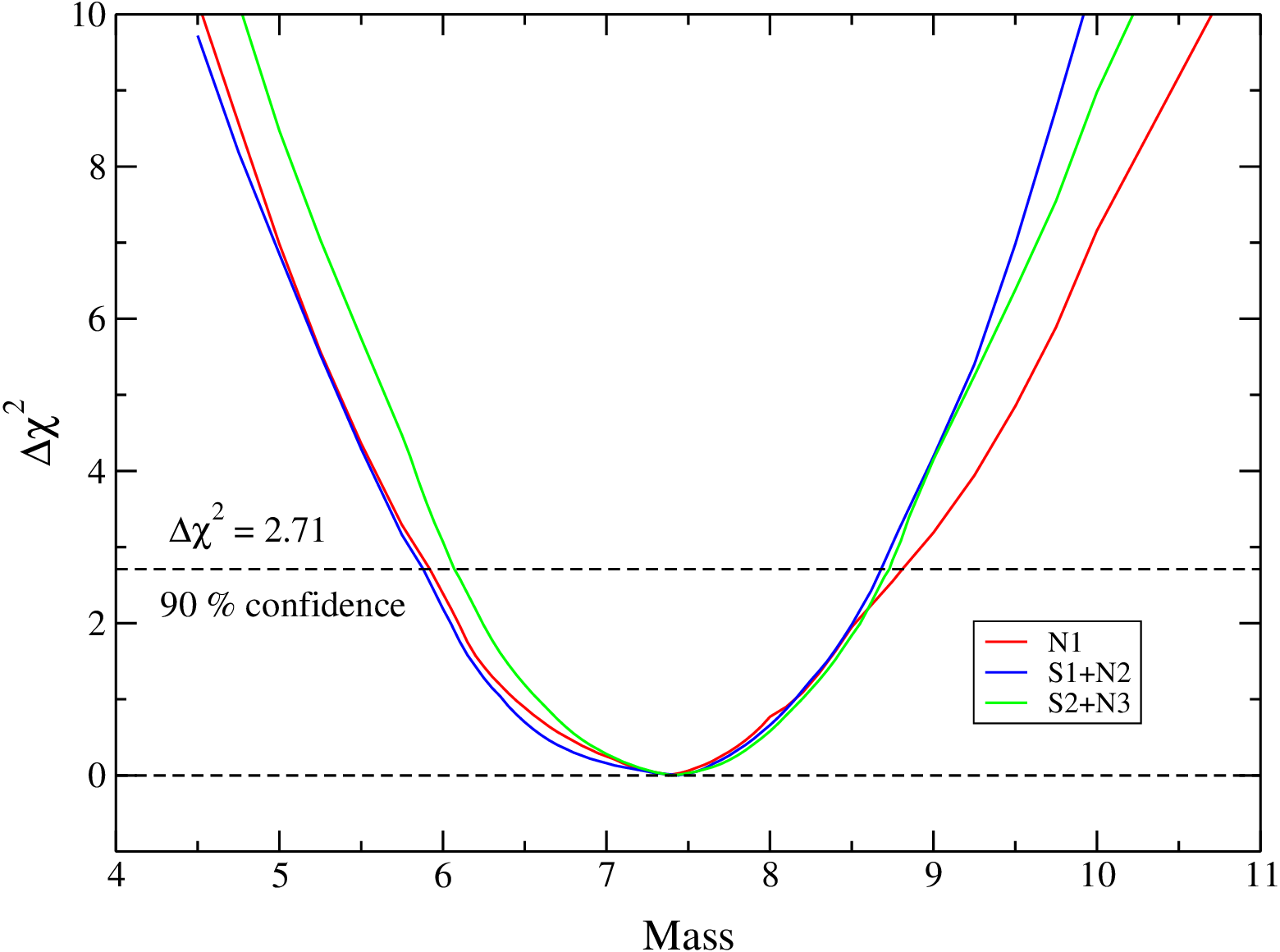}
\caption{The variation of $\Delta \chi^2$ is shown with the mass of the BH ($M_{BH}$) 
for three {\it NuSTAR} observations.}
\label{fig:mbh}
\end{figure}

\section{Discussion}

We studied MAXI~J1813-095 during its 2018 outburst using the data from {\it Swift}/XRT, {\it NICER}, and {\it NuSTAR} observatories in the energy range of $0.5-78$~keV. The {\it Swift}/XRT observed the source twelve times during the outburst. The $0.5-10$~keV {\it Swift}/XRT data were fitted with an absorbed power law model. Disk component was not required while fitting the spectra with the absorbed power law model. We also did not find any evidence of Fe K$\alpha$ line in the $0.5-10$~keV {\it Swift}/XRT data. {\it NICER} observed MAXI~J1813-095 at five epochs during the outburst. On contrary to the $0.5-10$~keV {\it Swift}/XRT spectra, the $0.5-10$~keV {\it NICER} spectra required a disk component along with the power law continuum. Superior spectral resolution of {\it NICER} over the {\it Swift}/XRT detected an additional spectral component, which {\it Swift}/XRT could not detect.

Interestingly, when we studied the $3-78$~keV {\it NuSTAR} (N1) spectra or the $0.5-78$~keV {\it Swift}/XRT+{\it NuSTAR} (N2+S1 and N3+S2) spectra, a disk component, an Fe K$\alpha$ emission line and reflection components were required along with the power law continuum. This suggests that the $0.5-10$~keV {\it Swift}/XRT or {\it NICER} spectra did not provide complete information on the source spectra. Moreover, the exposure time of each of the {\it NuSTAR} observation is long (on average $\sim20$ ks), while the {\it NICER} and {\it Swift}/XRT observations are of short exposure time ($\sim 1-2$ ks, see Table~\ref{tab:log}). Thus, the long exposure of {\it NuSTAR} and its broad-band coverage helped to detect additional spectral features in the source.

\subsection{Outburst Profile}

The 2018 outburst of MAXI~J1813-095 continued for $\sim 50$~days. The peak luminosity of the
source was observed on 2018 February 23, with $L_{peak} = 4.25 \times 10^{36} (d/6)^2$ ergs s$^{-1}$
in the $0.5-10$~keV energy band. However, one needs to calculate the luminosity in a broad energy range 
to extract a detailed information. We calculated bolometric luminosity of the source for three {\it NuSTAR}
observations from the unabsorbed flux in $0.1-100$~keV energy band. The bolometric luminosity of the source
was estimated to be $ L = 7.9-10.5 \times 10^{37}$ $(d/6)^2$ ergs s$^{-1}$. Thus, $L/L_{Edd} \sim 0.06-0.07$, for a BH with mass 7.4 $M_{\sun}$. Since, the observed mass accretion rate is $\sim 0.6$ $\dot{M}_{Edd}$, the accretion efficiency is very low during the 2018 outburst. 

During the entire outburst, the hard X-ray photons ($2-10$~keV range) dominate over the soft X-ray photons in $0.5-2$~keV range (see Fig.~\ref{fig:lc}). High HR was also observed during the outburst. We estimated thermal flux ($F_d$) and non-thermal flux ($F_{PL}$) from the {\tt diskbb} and {\tt powerlaw} model components, respectively, from the combined {\it Swift}/XRT and {\it NuSTAR} simultaneous spectral fitting in $0.5-78$~keV range. We find that the fraction of the thermal flux with respect to the non-thermal flux ($f_{th} = F_d / F_{PL}$) is less than 10\% in the $0.5-78$~keV range and less than 1\% in the $3-78$~keV energy range. The spectral analysis results (low $\Gamma$, high sub-Keplerian flow rate over the Keplerian flow rate, strong shock, etc.) indicate that the source was in the hard state during the {\it NuSTAR} observations. Thus, together with the spectral properties, the evolution of $0.5-2$~keV and $2-10$~keV fluxes, high HR, and HID, we infer that the source remained in the hard state during the entire outburst. Strong variabilities ($>20-30$ \% RMS) observed in the power density spectra also support this. 

The 2018 outburst of MAXI~J1813-095 can be considered as a `failed' outburst as the source
failed to make the state transition to softer spectral states. The observed HID of the source is similar to the HID of other sources during respective `failed' outbursts, where the HR do not change despite the change in the X-ray intensity \citep{Tetarenko2016}. A detailed study of 132 outbursts, \cite{Tetarenko2016} reported that the mean outburst duration for the `failed' outburst is about $\sim 290$ days. Although, many `failed' outbursts were observed to be as short as the 2018 outburst of MAXI J1813-095. For example, the outburst duration of 1998 outburst of XTE~J0421+560 \citep{Belloni1999} and  2011 outburst of Swift~J1357.2-0933 \citep{AP13} are 49 days and 76 days, respectively. In general, the failed outbursts are `faint' with peak luminosity ($L_{peak}) \lesssim 10^{36}$ ergs s$^{-1}$ (e.g. the 2000 outburst of XTE~J1118+480 \citep{DC2019}, the 2003 outburst of XTE~J1550-564 \citep{Stuner2005}), whereas the peak luminosity during complete outbursts, $L_{peak} \sim 10^{38}$ ergs s$^{-1}$ e.g. the 2009 outburst of XTE~J1752-223 \citep{Reis2011}, the 2017 outburst of MAXI~J1535-571 \citep{Stiele2018}, the 2019 outburst of MAXI~J1348-630 \citep{AJ20b}. The peak luminosity of MAXI~J1813-095 during the present outburst is $L_{peak} \sim 10^{36}$ ergs s$^{-1}$, which is consistent with other `failed' outbursts. Thus, MAXI~J1813-095 joined the ever-increasing list of `failed' outbursts  (e.g. the 2008 outburst of H~1743-322 \citep{Capitanio2009}, the 2011 \& 2012 outbursts of MAXI~J1836-194 \citep{AJ16,AJ20a}, the 2017 outburst of Swift~J1357.2-0933 \citep{SM19}; the 2017 outburst of GX~339-4 \citep{Garcia2019}.)

\subsection{Accretion Geometry}

In general, an outburst is triggered when the viscosity is suddenly enhanced at the outer edge of the disk \citep{Ebisawa1996}. The accreting matter loses angular momentum when the viscosity rises and rushes towards to the BH. The low viscous sub-Keplerian flow moves inward roughly in the free-fall time scale, whereas the Keplerian disk moves inward in the viscous time scale. If the viscosity is sufficiently high, the Keplerian disk moves closer to the black hole and cools the CENBOL and the source undergoes state transition \citep{Giri2012,SM17}. However, if the viscosity does not rise high enough, the Keplerian disk remains at a large distance from the black hole. Hence, the Keplerian disk cannot cool the CENBOL efficiently. As a result, the source does not enter to the softer spectral states. In the 2018 outburst of MAXI~J1813-095, it appears that the viscosity did not become high enough, and the source did not enter in the softer spectral states. Although, the Keplerian disk accretion rate was low, the continuous supply of the sub-Keplerian matter leads to increase of high energy flux, as well as the total flux, which leads to higher HR when the flux was high. This is not observed in regular outburst. The source entered the declining phase of the outburst when the viscosity is turned off. The shock moved outward as the accretion rates were decreased and the source entered to the quiescent state.

We did not observe any QPO in the PDS of the source. It is understood that the oscillation of the CENBOL or the Compton cloud produces the QPOs \citep{MSC96,Giri2012}. Sharp QPOs are produced when a strong shock oscillates and the resonance condition is satisfied \citep{MSC96,CMD15}. The resonance condition is satisfied when the cooling time of the post-shock matter matches with the infall time. On the other hand, a weak QPO is produced due to the non-satisfaction of the Rankine-Hugoniot condition or oscillation of the shock-less barrier or weak shock oscillation \citep{Ryu97}. A strong shock was formed during the 2018 outburst of MAXI~J1813-095. However, due to the low Keplerian disk accretion rate, and high sub-Keplerian halo rate, the cooling was inefficient. Thus, it is plausible that non-satisfaction of the resonance condition is behind of non-observation of QPOs. This is already reported in several sources that non-satisfaction of resonance condition is the reason behind non-observation of QPOs \citep{CMD15,AJ20a,AJ20b}.

In the first two {\it NuSTAR} observations (N1 \& N2), the source was observed in the decay phase, while the third observation (N3) was made in the brief re-brightening period. In the first two observations, we found that both accretion rates ($\dot{m}_d$ \& $\dot{m}_h$) decreased, though, in the third observation, the accretion rates marginally increased. The shock was found to move outward (93 $r_s$ to 113 $r_s$), although the shock strength remained stable ($\sim 2.80$). 

In TCAF, the normalization is a function of mass of the BH, distance and inclination angle of the system, and is given by $N_{tcaf} \sim (r_g^2/4 \pi d^2) \cos i$, where {\it d} is distance in 10~kpc and {\it i} is inclination angle. Thus ideally, one should find that the normalization is same for all the observations. However, there could be some fluctuations due to measurement errors. However, one could see a large deviation if jet is present \citep{AJ17,AJ18,DC2019}. In our analysis, we find $N_{tcaf} \sim 1.62, 1.65 ~\& 1.67$ for N1, N2 \& N3, respectively. This indicates that either there is no jet or a compact jet exists with very low outflow rate. Indeed, \cite{Russell18} observed a compact jet in the system. We calculated the mass outflow rate using Eqn.16 of \cite{C99}. The ratio of mass outflow rate ($\dot{M}_{out}$) to mass inflow rate (accretion rate, $\dot{M}_{in} = \dot{m}_d + \dot{m}_h$) given by, $r_{\dot{m}}= \frac{\dot{M}_{out}}{\dot{M}_{in}}=\frac{\theta_{out}}{\theta_{in}}\frac{R}{4} (\frac{R^2}{R-1})^{3/2}\exp(\frac{3}{2}-\frac{R^2}{R-1}),$ where, $\theta_{out}$ and $\theta_{in}$ are the solid angles subtended by the outflow and inflow, respectively. Using the TCAF model fitted $R$ and assuming $\theta_{out} \sim \theta_{in}$, we found that $\dot{M}_{out} \sim 0.003 \dot{M}_{in}$ during all three observations. Thus, the mass outflow rate is indeed very low and stable, hence, TCAF model normalization is constant.

The hard X-ray emission is reprocessed by the Keplerian accretion disk and contribute to the Fe K$\alpha$ emission line and reflection hump \citep{Guilbert1988,Lightman1988,Fabian89}. In general, the reflection hump is observed around $\sim 15-30$~keV. We studied the reflection feature of the spectra using the convolution model {\tt reflect}. We find that the reflection is weak with $R_{refl} = $ 0.15, 0.22 and 0.25 in N1, N2, and N3, respectively. Low accretion rate and far away location of the Keplerian disk are responsible for the weak reflection.

\subsection{Intrinsic properties of the system}

The mass of the black hole is a free parameter in TCAF model. Thus, we can estimate the mass of the black hole from the spectral analysis with the TCAF model. The masses of several BHs are already estimated from the spectral analysis with the TCAF model \citep{Molla16,DC16,Shang19}. The mass of MAXI~J1813-095 was unknown; thus we kept the mass free during the spectral analysis. The mass of the black hole was obtained as 7.38, 7.44 \& 7.40 $M_{\sun}$ in N1, N2 \& N3, respectively. We plot the variation of the mass with the $\Delta \chi^2$ in Fig.~\ref{fig:mbh}. Taking an average of three observations, we estimate the mass of the black hole as $7.41^{+1.47}_{-1.52}$ $M_{\sun}$, with 90\% confidence, or simply, $7.4 \pm 1.5$ $M_{\sun}$.

The Fe K$\alpha$ line is subjected to relativistic broadening if it is emitted from a region very close to the black hole \citep{Fabian89,Laor91}. In this work, we used relativistic model {\tt LAOR} to fit the {\it NuSTAR} data for broad iron K$\alpha$ emission line. In the process, we obtained the inner edge  of the accretion flow ($R_{in}$). Equating $R_{in}$ with the innermost stable orbit ($R_{isco}$), we can calculate the spin of the black hole. In this method, the spin of several black holes have been estimated \citep{Miller2004,Park2004,Reis2008,SM16}. For BHC MAXI~J1813-095, we obtained the inner edge of the accretion flow $R_{in}$ as $2.64^{+0.10}_{-0.06}$ $r_g$, $2.59^{+0.11}_{-0.08}$ $r_g$, and $2.58^{+0.11}_{-0.07}$ $r_g$, for N1, N2 and N3, respectively. This translates to the spin parameters ($a^*$) of the black hole as $0.74^{+0.02}_{-0.03}$, $0.75^{+0.02}_{-0.03}$, and $0.76^{+0.02}_{-0.03}$ for N1, N2 and N3, respectively. The accretion flow moves closer to the black hole in the soft state compared to the hard state. Since all the observations are taken in the hard state, $R_{in}$ would have moved closer in the soft state. Hence, the estimated spin parameters ($a^*$) only gives us the minimum value. Thus, we conclude the spin parameters of MAXI~J1813-095 as $a^* > 0.76$.

The strength of the reflection and the Fe line emission also depend on the inclination angle of the source. Thus, from the reflection and line emission, the inclination angle of the source can be constrained. We found the evidence for weak reflection in all three observations. From {\tt reflect} model fitted parameter, $\cos(incl)$ varied between 0.71 and 0.88, which translates to $\theta_{incl}$ between $28.36^{\circ}$ and $44.76^{\circ}$. From {\tt LAOR} model fitting, the $\theta_{incl}$ is between $31.88^{\circ}$ and $36.26^{\circ}$. Thus, the inclination angle of the source is $28^{\circ} - 45^{\circ}$. This low inclination angle naturally explains the low reflection of the source.

We also estimated the distance of the source from the unabsorbed flux and the Keplerian disk accretion rate. The source intrinsic luminosity, $L=\eta \dot{M} c^2 = 4 \pi d^2 F$, where $\eta$, $\dot{M}$, $c$, $d$, and $F$ are the accretion efficiency, mass accretion rate, light speed, source distance, and unabsorbed flux, respectively. The Eddington luminosity, $L_{Edd} = \eta \dot{M}_{Edd} c^2$, where $\dot{M}_{Edd}$ is the Eddington mass accretion rate. The TCAF model fitted mass accretion rate, $\dot{m} = \dot{M}/\dot{M}_{Edd}.$ From these three equations, we have $4 \pi d^2 F = \dot{m} L_{Edd}.$ From this equation, we estimated the distance from the three {\it NuSTAR} ~observations as 6.02 kpc, 5.83 kpc, and 6.25 kpc, for N1, N2, and N3, respectively. From this, the source distance is about $\sim 6$~kpc.

\section{Conclusions}
We studied the 2018 outburst of MAXI~J1813-095 using the data obtained from the {\it Swift}/XRT, {\it NICER} and {\it NuSTAR} observations. Our key findings are the following.

\begin{enumerate}
\item[1.] MAXI~J1813-095 remained in the hard state during the entire outburst. The source did not show state transition. This makes the outburst a `failed' outburst.

\item[2.] We studied power density spectra obtained from the $0.5-10$~keV {\it NICER} light curves. Strong 
variabilities were found with rms $\sim20-30$\%. We did not find any signature of QPO.

\item[3.] The $0.5-78$~keV {\it Swift}/XRT+{\it NuSTAR} spectra can be fitted well with the combined {\tt diskbb} 
and {\tt powerlaw} model. However, the fitting improved when we added a reflection component (modelled with
{\tt reflect} in {\tt XSPEC}). A presence of weak reflection was found in the spectra obtained from all three 
observations.

\item[4.] From the spectral analysis with the TCAF model, we extracted the accretion rates ($\dot{m}_d$ \& 
$\dot{m}_h$), size of the Compton cloud (i.e. the shock location, $X_s$), and the shock compression ratio 
($R$). We observed that the accretion rates decreased and the shock moved outward in the decay phase as the outburst progressed.

\item[5.] We found that the mass outflow rate which is very low and constant during our observations.

\item[6.] We estimated the mass of MAXI~J1813-095 as $7.4 \pm 1.5$ $M_{\sun}$. The distance of the source
is estimated to be $\sim 6$~kpc.

\item[7.] We estimated the spin parameter of the BH as $a^* > 0.76$. We also found that the inclination of the source 
is likely to be between $28^{\circ} - 45^{\circ}$.

\end{enumerate}

\begin{acknowledgements}
We thank the anonymous Reviewer for the constructive review which improved the clarity of the manuscript.
This research has made use of data and/or software provided by the High Energy Astrophysics Science Archive Research Center (HEASARC), which is a service of the Astrophysics Science Division at NASA/GSFC and the High Energy Astrophysics Division of the Smithsonian Astrophysical Observatory. This research has made use of the NuSTAR Data Analysis Software (NuSTARDAS) jointly developed by the ASI Science Data Center (ASDC, Italy) and the California Institute of Technology (Caltech, USA). This work was made use of XRT data supplied by the UK Swift Science Data Centre at the University of Leicester, UK. A.J. and N. K. acknowledges support from the research fellowship from Physical Research Laboratory, Ahmedabad, India, funded by the Department of Space, Government of India for this work. K.C. acknowledges support from DST/INSPIRE Fellowship (IF170233). R.B. acknowledges support from CSIR-UGC NET qualified UGC fellowship (June-2018, 527223). Research of S.K.C. and D.D. is supported in part by the Higher Education Dept. of the Govt. of West Bengal, India. S.K.C. and D.D. also acknowledge partial support from ISRO sponsored RESPOND project (ISRO/RES/2/418/17-18) fund. H.-K. C. is supported by MOST of Taiwan under grants MOST/106-2923-M-007-002-MY3 and MOST/108-2112-M-007-003. D.D. acknowledge support from DST/GITA sponsored India-Taiwan collaborative project (GITA/DST/TWN/P-76/2017) fund.

\end{acknowledgements}

\label{lastpage}

\end{document}